\documentclass[journal,letterpaper,twoside,twocolumn]{IEEEtran}
\usepackage{amsmath,amssymb,graphicx,psfrag,cite}
\usepackage[normalem]{ulem} 
\usepackage[usenames,dvipsnames]{color} 
\usepackage{balance}
\usepackage{mathrsfs}   
\usepackage{paralist}

\graphicspath{{figure/}}

\usepackage{color}

\definecolor{emphasis}{RGB}{204,0,204}
\renewcommand{\uline}[1]{{\color{emphasis}#1}}

  \renewcommand{\sout}[1]{} \renewcommand{\uline}[1]{#1} 

\title{Preferred Design of Hierarchical Distribution Matching} %
\author{Tsuyoshi~Yoshida,~\IEEEmembership{Member,~IEEE,} Magnus~Karlsson,~\IEEEmembership{Fellow,~OSA;~Senior~Member,~IEEE,} \\ and Erik~Agrell,~\IEEEmembership{Fellow,~IEEE} 
	\thanks{T.~Yoshida visited the Dept.~of Microtechnology and Nanoscience, Chalmers University of Technology, SE-412 96 G\"oteborg, Sweden, and is now with Information Technology R\&D Center, Mitsubishi Electric Corporation, Kamakura, 247-8501, Japan. He also belongs to Graduate School of Engineering, Osaka University, Suita, 505-0871, Japan (e-mail:  Yoshida.Tsuyoshi@ah.MitsubishiElectric.co.jp).}
	\thanks{M.~Karlsson is with the Dept.~of Microtechnology and Nanoscience and E.~Agrell is with the Dept.~of Electrical Engineering, both at Chalmers University of Technology.}
	\thanks{Patent application of hierarchical distribution matching has done on Aug. 7, 2018 (application number: PCT/JP2018/029578).}
	\thanks{This work was partly supported by ``Massively Parallel and Sliced Optical Network (MAPLE),'' the Commissioned Research of National Institute of Information and Communications Technology (NICT), Japan.}
}%
\begin{document}
\maketitle

\begin{abstract}
Distribution matching and dematching (DM/invDM) are key functions in probabilistic shaping (PS). Recently techniques for low complexity implementation of DM/invDM have been well studied. Our previously proposed hierarchical DM (HiDM) is one of the good candidates, with capacity-approaching performance with reasonable hardware resources. Though we explained the recipe of HiDM construction with a small example having a short DM word length, there might still be difficulties to expand it to longer DM word lengths. To improve the reproducibility of our work, this paper explains the key parameters in an HiDM having a DM word length more than 100 symbols.
\end{abstract}


\section{Introduction}
\label{sec:intro}
Constellation shaping has been deeply investigated over several decades to approach the Shannon capacity over the Gaussian channel. There are two shaping schemes, geometric shaping \cite{forney_1989_jsac} and probabilistic shaping (PS) \cite{calderbank_1990_tit,kschischang_1993_tit}. 
Coded modulation techniques have received much interest in the optical fiber communication field \cite{agrell_2009_jlt} after the deployment of coherent detection with digital signal processing \cite{roberts_2009_jlt}.
Probabilistic amplitude shaping (PAS) \cite{bocherer_2015_tcom} proposed in the communication theory field significantly improve the implementation possibility of PS by using reverse concatenation \cite{bliss_1981_ibm,fan_1999_globecom,djordjevic_2006_jlt}, which means forward error correction (FEC) inside the shaping DM/invDM. The PAS scheme was early examined in optical fiber communications \cite{buchali_2016_jlt} and gave a significant impact in the community.

Fiber-optic communication channels with optical amplifiers are suitable target applications of PS. The first reason is the existence of the linear optical amplifier. When we shape the optical signal, we reduce the average optical power inside an optical modulator, but the power will soon be recovered by optical amplifiers, which gives an almost linear gain without waveform degradation. The second reason is the channel stability because of the confined waveguide (fiber) transmission. Though some signal interferences would remain, a fiber-optic communication channel can be approximated by a Gaussian channel with an average power constraint.

The shaping encoding and decoding functions for PS are called distribution matching (DM) and distribution dematching (invDM), resp. Their high performance and low complexity implementation is one of the hot topics in this field \cite{ramabadran_1990_tcom,schulte_2016_tit,cho_2016_ecoc,bocherer_2017_arxiv_spg,yoshida_2017_ecoc_spg,bocherer_2017_ecoc,pikus_2017_comlet,gultekin_2017_pimrc,
yoshida_2018_ofc_spg,steiner_2018_ciss,gultekin_2018_sitb,gultekin_2018_isit,schulte_2018_arxiv,yoshida_2018_ecoc,
fehenberger_2019_tcom,yoshida_2019_jlt,cho_2019_arxiv,koganei_2019_ofc,yoshida_2019_ofc,cho_2019_ofc}. Among the DMs, our previously proposed \emph{hierarchical DM (HiDM)} having a unique tree structure of look-up tables (LUTs) shows good performance, reasonable complexity leading to small power consumption for the implementation, high throughput, and small error rate increase in the invDM processing.
The paper \cite{yoshida_2019_jlt} explained the recipes to configure the LUT tree and to choose the LUT contents for the HiDM by using a small example. However, it would be informative and help others reproduce our results to explain detailed parameters also in the case of long DM word lengths. Thus this paper discloses a set of exemplified LUT interface parameters to design a HiDM having more than 100 quadrature amplitude modulated (QAM) symbols.

\begin{figure}[tb]
	\begin{center}
		\setlength{\unitlength}{.6mm} %
		\scriptsize
		\vspace{-0.1cm}
		\includegraphics[scale=0.48]{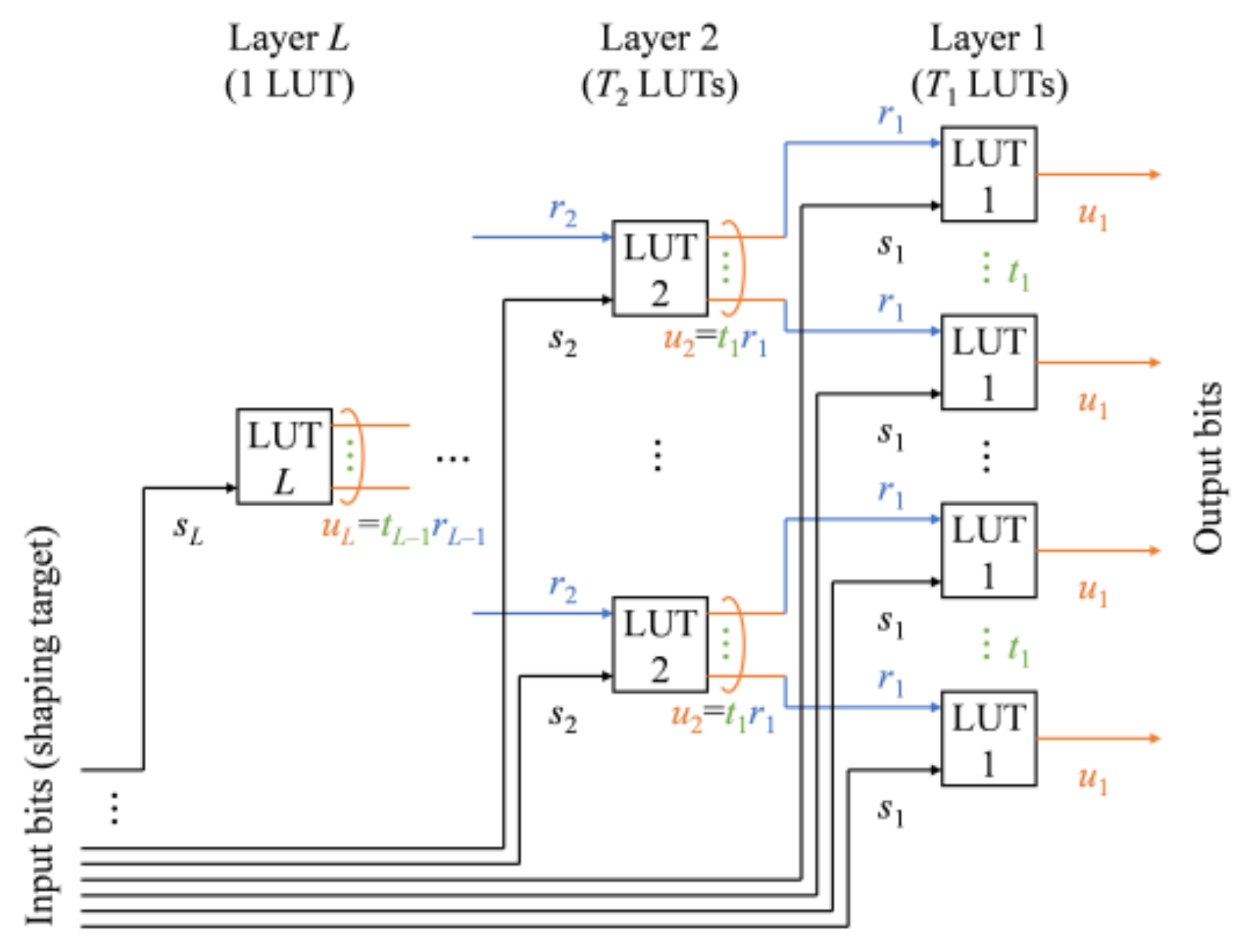} \\
		\vspace{-0.3cm}		
		\caption{Schematic of hierarchical DM.}
		\label{fig:dm}
	\end{center}
	\vspace{-0.4cm}
\end{figure}

\section{Design and evaluation of hierarchical DM (HiDM)}
\label{sec:system}

Fig.~\ref{fig:dm} shows the schematic of the HiDM \cite{yoshida_2019_jlt}. For simplicity, we exclude non-shaped bits from the figure. 
The parameters in reverse concatenation PS and HiDM are shown in Tabs.~\ref{tab:param_rcps} and \ref{tab:key_param}, resp.
There are $N_{\text{in}}$ input bits as a shaping target. They are separately input to LUTs, hierarchically placed in $L$ layers. In the top layer, an LUT receives $s_{L}$ bits and outputs $u_{L}=t_{L-1} r_{L-1}$ bits. The $r_{L-1}$ bits are input to an LUT in layer $L-1$, and the $t_{L -1}$ LUTs in layer $L-1$ are connected to an LUT in layer $L$. In layer $\ell$ ($\neq L$ or 1), each LUT receives $r_{\ell}$ bits from layer $\ell +1$ and $s_{\ell}$ bits from the input of the DM as information bits. Totally $v_{\ell} = r_{\ell} + s_{\ell}$ bits are converted into $u_{\ell}=t_{\ell -1} r_{\ell -1}$ bits, which are fed to layer $\ell - 1$. The number of LUTs in layer $\ell$ is denoted as $T_{\ell}$. In layer 1 each LUT receives $r_1$ bits from layer 2 and $s_1$ bits from the input of the DM. Totally $v_1 = r_1 + s_1$ bits are converted into $u_1$ bits, which corresponds to $u_1 / m^{\text{sb}}$ QAM symbols. The number of DM output bits and QAM symbols are $m^{\text{sb}}N_{\text{s}} = T_1 u_1 = \Pi_{\ell = 1}^{L-1} t_{\ell} u_1$ and $N_{\text{s}}/2$, resp.

In an example for PS-256-QAM generation \cite{yoshida_2019_jlt}, the total number of bits per complex symbol $m$ is 8, and both the sign bits (the most significant bits) and the least significant bits are not shaped. Only the second and third significant bits are shaped in each dimension, so that $m^{\text{sb}}=4$.
Tab.~\ref{tab:eg_param} exemplifies the parameters used.
The number of DM input bits per DM word $\sum_{\ell =1}^{L} T_{\ell} s_{\ell}$ is 507, and the number of DM output bits per DM word $m^{\text{sb}} N_{\text{s}} = T_1 u_1$ is 640.
Thus the maximum spectral efficiency per 2D symbol at an FEC code rate of 1 is $\beta = 2 (2+507/320)$ bit per channel use (bpcu). The entropy of a 2D symbol $2H(X)$ will be larger than $\beta$, where $X$ denotes a pulse amplitude modulation (PAM) symbol.

The values of $T_{\ell}$, $v_{\ell}$, and $u_{\ell}$ determine the accumulated size of the LUTs, i.e., $\sum_{\ell = 1}^L T_{\ell} 2^{v_{\ell}} u_{\ell}$ for DM. When we employ a simple mirror structure for the invDM, its size will be\footnote{Potentially there would be techniques to reduce the LUT size under the same performance.} $\sum_{\ell = 1}^L T_{\ell} 2^{u_{\ell}} v_{\ell}$. Thus, practically we would have constraints on the values of $v_{\ell}$ and $u_{\ell}$, which depend on what hardware resource use is acceptable. Under such a constraint, a binary tree structure $t_{\ell} = 2$ gives the best shaping performance.

\begin{table}[t]
	\caption{Parameters in reverse concatenation PS systems.}
	\label{tab:param_rcps}
	\vspace{-0.4cm}
	\begin{center}
	\begin{tabular}{cl}
		\hline
		Notation & Description \\
		\hline\hline
		$m$ & number of bits per QAM symbol \\
		$m^{\text{sb}}$ & number of shaped bits per QAM symbol \\
		$N_{\text{s}}$ & number of PAM symbols per DM word \\
		$N_{\text{in}}$ & number of information bits per DM word \\
		\hline
	\end{tabular}
	\end{center}
\end{table}

\begin{table}[t]
	\caption{Key parameters in hierarchical DM.}
	\label{tab:key_param}
	\vspace{-0.4cm}
	\begin{center}
	\begin{tabular}{cl}
		\hline
		Notation & Description \\
		\hline\hline
		$\ell$ & layer index \\
		$L$ & number of layers \\
		$t_{\ell}$ & number of LUTs in layer $\ell$ connected to an LUT \\
		           & in layer $\ell +1$ \\
		$T_{\ell}$ & number of LUTs in layer $\ell$ \\
		$r_{\ell}$ & number of input bits in an LUT in layer $\ell$ from layer $\ell +1$ \\ 
		$s_{\ell}$ & number of input bits in an LUT in layer $\ell$ as information bits \\
		$v_{\ell}$ & number of input bits in an LUT in layer $\ell$ ($r_{\ell}+s_{\ell}$) \\
		$u_{\ell}$ & number of output bits in an LUT in layer $\ell$ \\
		            & ($mN_{\text{s}}/2$ if $\ell = 1$ or $2r_{\ell -1}$ else) \\
		\hline
	\end{tabular}
	\end{center}
\end{table}

\begin{table}[t]
	\caption{Chosen parameters used in \cite[Tab.~IV, Fig.~4]{yoshida_2019_jlt}.}
	\label{tab:eg_param}
	\vspace{-0.4cm}
	\begin{center}
	\begin{tabular}{c|cccccc}
		\hline
		$\ell$ & $t_{\ell}$ & $T_{\ell}$ & $r_{\ell}$ & $s_{\ell}$ & $v_{\ell}$ & $u_{\ell}$ \\
		\hline\hline
		7 &    & 1  &   & 5 & 5 & 12 \\ 
		6 & 2 & 2  & 6 & 5 & 11 & 12 \\   
		5 & 2 & 4  & 6 & 5 & 11 & 12 \\ 
		4 & 2 & 8  & 6 & 5 & 11 & 12 \\ 
		3 & 2 & 16 & 6 & 5 & 11 & 12 \\
		2 & 2 & 32 & 6 & 5 & 11 & 12 \\
		1 & 2 & 64 & 6 & 3 & 9 & 10 \\
		\hline
	\end{tabular}
	\end{center}
\end{table}

We generated PS-256-QAM signals having a DM word length of 320 16-PAM symbols by employing constant composition DM (CCDM) \cite{schulte_2016_tit} and HiDM \cite{yoshida_2019_jlt}. In Tab.~\ref{tab:stat_spg} \cite{yoshida_2019_jlt}, we summarize the statistics, i.e., the probability mass function $P_X$, average QAM symbol energy $E$, QAM symbol entropy $2 H(X)$, 
maximum spectral efficiency $\beta$, 
rate loss $R_{\text{loss}} = 2H(X) - \beta$, and constellation gain $G = (2^{\beta}-1) d_{\text{min}}^2 / (6E)$, where $d_{\text{min}}$ denotes minimum Euclidean distance, of the generated signals. The rate loss of a QAM symbol were 0.07 and 0.08 bpcu for CCDM and HiDM, resp. In each case, the constellation gain was more than 1 dB, whose gap from the ideal Maxwell--Boltzmann (MB) distribution was within 0.4 dB even though we did not shape the least significant bit.

\begin{table}
	\caption{Statistics of the shaped signals \cite{yoshida_2019_jlt}.}
	\label{tab:stat_spg}
	\vspace{-0.4cm}
	\begin{center}
	\begin{tabular}{cccc}
		\hline
		& CCDM & HiDM & MB \\
		\hline\hline
		$N_{\text{s}}$ (PAM symbol) & 320 & 320 & -- \\
		$P_{|X|}(1)$ & 0.2453 & 0.2376 & 0.2628 \\
		$P_{|X|}(3)$ & 0.2453 & 0.2376 & 0.2355 \\
		$P_{|X|}(5)$ & 0.1625 & 0.1684 & 0.1891 \\
		$P_{|X|}(7)$ & 0.1625 & 0.1684 & 0.1360 \\
		$P_{|X|}(9)$ & 0.0719 & 0.0757 & 0.0877 \\
		$P_{|X|}(11)$ & 0.0719 & 0.0757 & 0.0506 \\
		$P_{|X|}(13)$ & 0.0203 & 0.0183 & 0.0262 \\
		$P_{|X|}(15)$ & 0.0203 & 0.0183 & 0.0121 \\
        	$E$ & 74.00 & 74.70 & 68.31 \\
        	$2H(X)$ (bpcu) & 7.242 & 7.252 & 7.169 \\
        	$\beta$ at $R_{\text{c}} = 1$ (bpcu) & 7.169 & 7.169 & 7.169 \\
        	$R_{\text{loss}}$ (bpcu) & 0.073 & 0.083 & 0 \\
		$G$ (dB) & 1.097 & 1.056 & 1.444 \\
		\hline
	\end{tabular}
	\end{center}
\end{table}

\section{Summary}
\label{sec:cncl}
In this contribution we explained the design of HiDM to improve reproducibility also in the case of long symbol sequences. The 7-layer configuration realizes a DM word length of 160 256-QAM symbols. The resulting energy gap from the ideal Maxwell--Boltzmann distribution is less than 0.4 dB, while keeping four bits per QAM symbol uniformly distributed (non-shaped). As shown in \cite{yoshida_2019_ofc}, this hierarchical DM is useful also for joint source--channel coding. This realizes simultaneous data compression and probabilistic shaping, which can further reduce the required signal-to-noise ratio 
or 
system power consumption in future optical networks.

\section*{Acknowledgments}
We thank Koji Igarashi of Osaka University for fruitful discussions.



\end{document}